\documentclass[%
 reprint,
superscriptaddress,
 amsmath,amssymb,
 aps,
]{revtex4-2}

\usepackage{graphicx}
\usepackage{dcolumn}
\usepackage{bm}

\usepackage{hyperref}
\usepackage[none]{hyphenat}
\usepackage{color}

\newcommand{\cts}{Co$_{1/4}$TaSe$_2$}
\newcommand{\C}{$^\circ$C}
\newcommand{\muB}{$\mu_\mathrm{B}$}
\newcommand{\mueff}{$\mu_\mathrm{eff}$}
\newcommand{\TN}{$T_N$}
\newcommand{\DT}{$\Theta_\mathrm{D}$}
\newcommand{\ET}{$\Theta_\mathrm{E}$}

\begin{document}

\preprint{APS/123-QED}

\title{Itinerant A-type Antiferromagnetic Order in \cts\ }

\author{H. Cein Mandujano}
\affiliation{Department of Chemistry and Biochemistry, University of Maryland, College Park, Maryland 20742, USA}
\affiliation{Maryland Quantum Materials Center, Department of Physics, University of Maryland, College Park, Maryland 20742, USA}
 
\author{Gicela Saucedo Salas}
\affiliation{NIST Center for Neutron Research, National Institute of Standards and Technology, 100 Bureau Drive, Gaithersburg, Maryland 20899, USA}
\affiliation{Maryland Quantum Materials Center, Department of Physics, University of Maryland, College Park, Maryland 20742, USA}%

\author{Tianyu Li}
\affiliation{Department of Chemistry and Biochemistry, University of Maryland, College Park, Maryland 20742, USA}

\author{Peter Y. Zavalij}
\affiliation{Department of Chemistry and Biochemistry, University of Maryland, College Park, Maryland 20742, USA}

\author{Alicia Manj\'on-Sanz}
\affiliation{Neutron Scattering Division, Oak Ridge
National Laboratory, Oak Ridge, Tennessee 37831, United
States}

\author{Nicholas P. Butch}
\affiliation{NIST Center for Neutron Research, National Institute of Standards and Technology, 100 Bureau Drive, Gaithersburg, Maryland 20899, USA}
\affiliation{Maryland Quantum Materials Center, Department of Physics, University of Maryland, College Park, Maryland 20742, USA}

\author{Efrain E. Rodriguez}
\email{efrain@umd.edu}
\affiliation{Department of Chemistry and Biochemistry, University of Maryland, College Park, Maryland 20742, USA}
\affiliation{Maryland Quantum Materials Center, Department of Physics, University of Maryland, College Park, Maryland 20742, USA}


\date{\today}

\begin{abstract}

We present the magnetic behavior and resulting transport properties of TaSe$_2$ when intercalated by magnetically active cobalt cations. Acting as the layered host, TaSe$_2$ is a transition metal dichalcogenide (TMD) that adopts the 2H-polytype. We find through our single crystal and powder diffraction studies that we can prepare the stoichiometry \cts, which crystallizes in the centrosymmetric space group $P6_3/mmc$. From magnetic susceptibility and x-ray photoelectron spectroscopy measurements, we find a transition consistent with antiferromagnetic order below the temperature \TN\ = 173 K and Co$^{2+}$ in the high-spin state. Neutron powder diffraction and specific heat measurements, however, point to a much smaller than anticipated ordered moment in this sample. From the neutron results, the magnetic structure can be described as an A-type antiferromagnet with an ordered moment size of 1.35(11) $\mu_B$ per Co cation. The direction of the moments are all long the c-axis, which is consistent with the magnetization and susceptibility studies showing this direction to be the easy axis. Interestingly, we find that a weak and subtle ferromagnetic component appears to exist along the $ab$-plane of the \cts\ crystals. We place the results of this work in the context of other magnetic-ion intercalated TMDs, especially those of Ta and Nb.

\end{abstract}

\maketitle


\section{\label{sec:level1} Introduction}

The discovery of ferromagnetism in two-dimensional (2D) materials has brought increasing interest to the subject of itinerant 2D magnetism.\cite{Gong2017} The nature of such lower-dimensional magnetism can arise from spin polarization of the delocalized electrons in these strongly anisotropic systems.\cite{Fei2018} Transition metal dichalcogenides (TMDs) constitute an important and broad category of van der Waals (vdW) materials making them a rich platform to explore such 2D magnetism.\cite{Chhowalla2013} Particularly, niobium- and tantalum-based TMDs are known metallic systems that can host other physical phenomena such as charge density wave (CDW) instabilities.\cite{Rossnagel2005, Laverock2013} Notably, commensurate and incommensurate CDWs as well as superconductivity can manifest in the metallic 2H-TaSe$_2$ polytype.\cite{Koyama1987} Using these inherently metallic TMD hosts, one can then design ferro- or antiferromagnetic vdW systems through insertion of magnetic guest species and explore how such magnetism interacts with CDWs or even superconductivity. In this article, we explore 2H-TaSe$_2$ as a host for exploring itinerant magnetism via magnetic cation intercalation.

The ease with which magnetically active metal ions can be inserted into TMD hosts has initiated a renewed surge in studying such intercalated systems.\cite{Hatanaka2023}. Typically, the magnetic ion is a member of the first-row transition metal series such as V, Cr, Mn, Fe, Co, and Ni. The guest ion stabilizes a 3$d$-electronic band that interacts with the itinerant electronic bands of the TMD host.\cite{PARKIN1980, Xie2022, Naik2022} Depending on the combination of the guest magnetic ion and TMD host, the symmetry of the overall system can also be altered, \textit{e.g.}, from centrosymmetric to noncentrosymmetric. The alteration of symmetry, both chemical and magnetic, could lead to emergent properties such as the anomalous Hall effect\cite{Zheng2021} or anisotropic magnetoresistance.\cite{Liu2021_Mn} For example, a proposed type of magnetic order known as altermagnetism has been postulated in a vanadium-intercalated TMD.\cite{Smejkal2022} Unlike traditional antiferromagnets, altermagnets break time-reversal symmetry in their spin-resolved electronic band structure despite also having anti-aligned magnetic moments in real space. Since TMDs could offer a platform for exploring altermagnetism, it is imperative that metallic TMDS with long-range magnetic order be explored.


The first criterion in these studies of itinerant magnetism is to choose a TMD host that is already metallic in its pristine state. Typically, the platforms for these studies consist of metals from groups 5 (\textit{e.g.}, Nb and Ta) and 7 (\textit{e.g.}, Tc, Re) of the periodic table.\cite{Chhowalla2013} The nature of the resulting magnetic order, whether ferro- or antiferromagnetic, must therefore arise from the behavior of the electrons near the Fermi level. For example, itinerant ferromagnetism is understood to arise from conduction band-splitting that leads spin imbalance,\cite{Moriya1984} and the emerging moments can be explained by the Stoner model\cite{Santiago2017}. In the case of itinerant antiferromagnetism (AFM), a variety of structures of structures have been observed including spin density waves, exotic spin textures, spin spirals, skyrmions, and helimagnets.\cite{Wu2018, Rodriguez2011, Gubkin2016, Cao2020, Yasui2020, Karna2019}

\begin{figure}[t]
\includegraphics[scale=0.45]{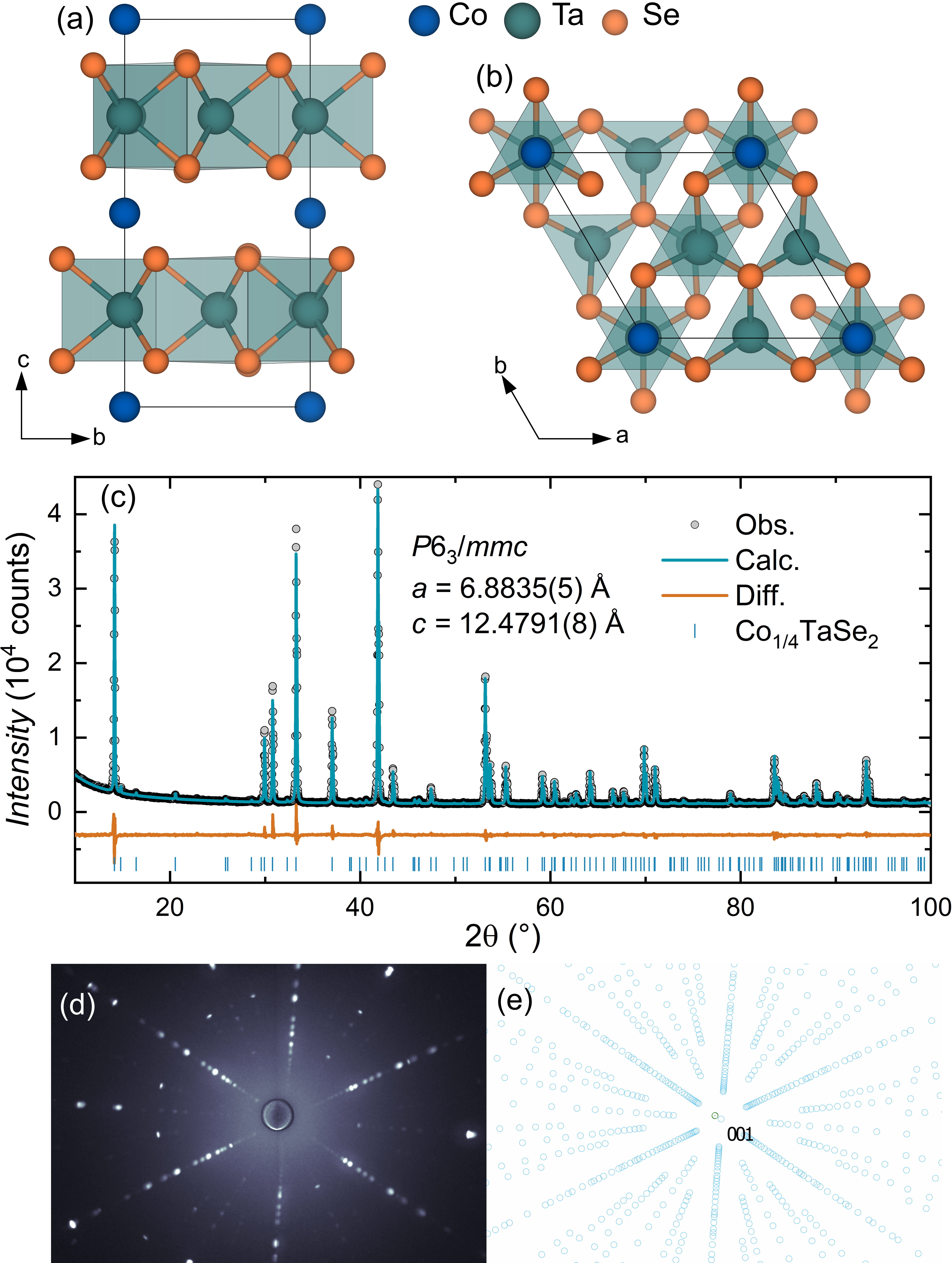}
\caption{\label{fig_xrd} Crystal structure of \cts\ and representative diffraction patterns. The layered 2H-polytype of \cts\ viewed along the $a$-axis (a) and $c$-axis (b). (c) Rietveld refinement of the crystal structure with pXRD using Cu K-$\alpha$ radiation. The residual was 4.86 \%, and the vertical tickmarks index the hexagonal cell. Cobalt sits in the vdW gap in the 2$a$ Wyckoff site to form a triangular sublattice. (d) Experimental Laue diffraction image and (e) simulated Laue diffraction showing the six-fold symmetry viewed along the $c$-direction.}
\end{figure}

In this work, we grow single crystals of Co$_{0.25}$TaSe$_2$ (Figure \ref{fig_xrd}(a) and (b)) and characterize their bulk physical properties including any anisotropy in their magnetic behavior. To the best of our knowledge, only Fe and Ni have been reported as magnetic ion intercalants into TaSe$_2$.\cite{ Maksimovic2022, Nosova2024} Both  Fe$_{0.25}$TaSe$_2$ and Ni$_{0.25}$TaSe$_2$ are known ferromagnets, albeit with different characteristics in their magnetic hysteresis. Much more explored are the sulfide-based TMDs for magnetic ion intercalation. For example, the sulfide host $M_x$TaS$_2$ can take up high ratios of metal intercalation, up to $x=\frac{1}{3}$, and a variety of phenomena are associated with the amount of $x$ and the $d-$electron count of $M$ in $M_x$TaS$_2$ ($M =$ V, Cr, Mn, Fe, Co, and Ni). Whereas the$M_x$TaS$_2$ exhibits helimagnetism for $M=$ Cr,\cite{Obeyesekera2021} it adopts several collinear AFM structures for $M=$ V, Co, and, Ni.\cite{Lu2020, Liu2021, An2023} Meanwhile, for the case of $M=$ Mn and Fe, they are all ferromagnets.\cite{Liu2021, Mangelsen2020} Given the richness of the magnetic phase diagrams found in the sulfide TMDs, it is worth exploring the selenide counterparts. The present work would constitute the first report into the effects of Co-intercalation into TaSe$_2$.

\section{Experimental Methods}


\subsection{Sample preparation and crystal growth}

Polycrystalline powder samples were made by vacuum sealing Co, Ta, and Se in 1:4:8 molar ratio in a fused silica ampule at $<10$ mTorr. Each preparation ranged from 1 g to 4 g of sample per ampule. The ampule with the premixed reagents was heated up to 700\C\ for 1 hour before increasing the temperature further to 900 \C\ where it dwelled for 5 days. After that, the samples were furnace cooled to room temperature. The resulting powders were recovered in air, reground in an agate mortar and pestle, and reheated to 900\C\ for 5 days to improve homogeneity.

Single crystal samples of \cts\ were obtained by chemical vapor transport (CVT) using pre-reacted powder samples and 4 mg of iodine per mL of the volume of the reaction ampule. However, due to iodine's affinity for cobalt, excess cobalt was added in the solid-state synthesis before CVT. The charge powder was thus prepared with a 1:1:2 molar ratio of Co:Ta:Se. Iodine and charge powder were loaded into a 12.5 cm-long fused silica tube with an inner diameter of 14 \color{black}mm \color{black} and outer diameter of 16 \color{black}mm\color{black}. The ampule was torch-sealed under partial vacuum and placed in a horizontal tube furnace to ensure that both precursors were at the center of a single-zone tube furnace. The hot end (furnace center) of the ampule was set to to 940 \C\ and the cool end was measured to be 880 \C\ . The temperature profile of the single-zone furnace was given by the natural gradient of the furnace.

Single crystals of \cts\ grew at the hot end of the ampule. Once recovered, they were washed with ethanol to remove residual iodine and any surface impurities. We had initially assumed the cool end of the ampule to be the growth zone, but for this system, the hot end is the actual growth zone. This observation would entail that the reaction of the transport agent with the precursor powder is exothermic.

\subsection{Sample characterization and electronic structure calculations}

For structure determination, a suitable single crystal of \cts\ was selected and measured on a Bruker D8Venture with PhotonIII diffractometer. 
The crystal was kept at 250(2) K during data collection. 
The integral intensity was correct for absorption using SADABS software\cite{Krause2015} using the integration method. 
The structure was solved with the ShelXT program and refined with the ShelXL\cite{Sheldrick2015} program using least-square minimization. 
No restraints were used in the structural refinement. More details on the structure determination by single crystal methods can be found in Table \ref{tab_scxrd}.

Neutron powder diffraction data was collected at the POWGEN instrument\cite{POWGEN_Huq2011} located at the Spallation Neutron Source in Oak Ridge National Laboratory.  Experiments at 200 K, 120 K, and 10 K were performed on approximately 8 g of powder sample loaded into standard vanadium cans. 
Two different time-of-flight frames were chosen to cover 0.5-12.5 \AA\ and 1-20.5 \AA \, in $d-$spacing, respectively. Both powder X-ray and neutron diffraction data were analyzed using the Rietveld method with GSAS-II\cite{gsas} software.

Powder X-ray diffraction was measured in reflection geometry on a Bruker D8 instrument with Cu K$\alpha$ wavelength and a LynxEye detector. 

The elemental composition of the single crystal samples was determined using energy dispersive X-ray spectroscopy (EDS). The EDS measurements were carried out on multiple areas of several crystals from the same batch in a Hitachi SU-70 Schottky field emission gun scanning electron microscope (FEG SEM). Further composition analysis was corroborated via X-ray photoelectron spectroscopy (XPS) in a Kratos Axis Supra X-ray Photoelectron Spectrometer with an Al anode source.  XPS analysis was performed using the CASAXPS\cite{CASAXPS}, and the binding energies obtained are relative to the Fermi level.

Magnetization isotherms, as well as temperature-dependent zero-field, cooled (ZFC) and field-cooled (FC) DC magnetic susceptibility measurements were performed using a Quantum Design Magnetic Property Measurement System (MPMS). Resistivity and magnetoresistance measurements were done in a 14 T Quantum Design Physical Property Measurement System (PPMS-14). Specific heat was measured in a Quantum Design Dynacool PPMS upon cooling using the two-tau method.

Self-consistent field and density of states calculations were performed using the Quantum ESPRESSO software package\cite{QE2009} in an 8$\times$8$\times$2 K-point mesh and cutoff energy of 680 eV. Projector-augmented wave (PAW)-type pseudopotentials were used with the generalized gradient approximation (GGA)+U.

\section{Results and Discussion}

\subsection{Structure solution and refinement}

\begin{figure}[t]
\includegraphics[scale=0.268]{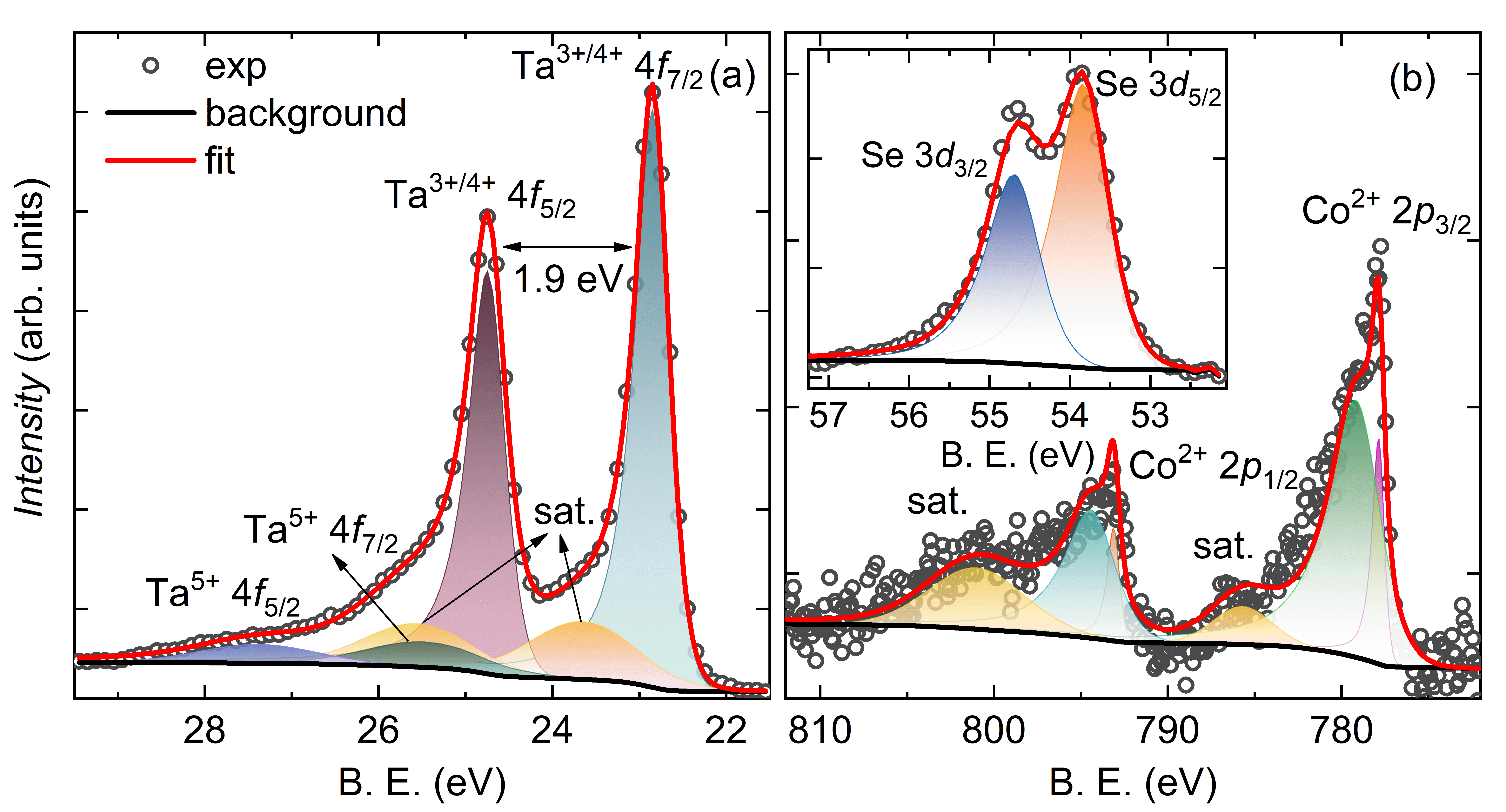}
\caption{\label{fig_xps} X-ray photoelectron spectroscopy analysis of \cts\ showing the spin-orbit energy splitting of (a) Ta $f$, (b) Co $p$, and (inset) Se $d$ doublets.}
\end{figure}

\begin{table}[!b]
\caption{\label{tab_scxrd} Single-crystal XRD structural solution results for \cts.}
\begin{ruledtabular}
\begin{tabular}{@{} cc @{\hspace{5mm}} ll @{}}
\multicolumn{2}{c}{Cryst system} & \multicolumn{2}{c}{Hexagonal} \\
\multicolumn{2}{c}{Space group} & \multicolumn{2}{c}{$P$6$_3$/$mmc$} \\
\multicolumn{2}{c}{$a$} & \multicolumn{2}{c}{6.8828(1) \AA} \\
\multicolumn{2}{c}{$c$} & \multicolumn{2}{c}{12.4535(3) \AA} \\
\multicolumn{2}{c}{Volume} & \multicolumn{2}{c}{510.92(2) \AA$^{3}$} \\
\multicolumn{2}{c}{$Z$} & \multicolumn{2}{c}{8} \\ 
\multicolumn{2}{c}{Calculated density} & \multicolumn{2}{c}{9.194 g/cm$^3$} \\ 
\multicolumn{2}{c}{F(000)} & \multicolumn{2}{c}{1182.0} \\
\multicolumn{2}{c}{Crystal size (mm)} & \multicolumn{2}{c}{0.24 $\times$ 0.15 $\times$ 0.015} \\
\multicolumn{2}{c}{Radiation} & \multicolumn{2}{c}{Mo$K_\alpha$($\lambda$ = 0.71073 \AA)} \\
\multicolumn{2}{c}{2$\Theta$ range} & \multicolumn{2}{c}{6.544$^{\circ}$ to 70.034$^{\circ}$} \\
\multicolumn{2}{c}{Refl collected} & \multicolumn{2}{c}{20723} \\
\multicolumn{2}{c}{Independent refl} & \multicolumn{2}{c}{470} \\
\multicolumn{2}{c}{G.O.F.on F$^2$} & \multicolumn{2}{c}{1.349} \\
\multicolumn{2}{c}{Final R indexes} & R$_1$ & 0.0164 \\
\multicolumn{2}{c}{[I$\geq$2$\sigma$ (I)]} & wR$_2$ & 0.0359 \\
\multicolumn{2}{c}{Final R indexes} & R$_1$ & 0.0169 \\
\multicolumn{2}{c}{[all data]} & wR$_2$ & 0.0360 \\
\multicolumn{2}{c}{Temperature} & \multicolumn{2}{c}{250(2) K} \\
\end{tabular}
\end{ruledtabular}
\end{table}

\begin{table*}[t]
\caption{\label{tab_scxrd_r} Structural parameters for \cts\ obtained from single-crystal XRD results at 250 K.}
\begin{ruledtabular}
\begin{tabular}{cccccccccccc}
 Atom&  Wyc. Pos.   &$x/a$  &$y/b$ &$z/c$   & Occ.  & $U_{11}$ & $U_{22}$ & $U_{33}$ & $U_{23}$ & $U_{13}$ & $U_{12}$\\ \hline
 Co &2$a$           &0	   & 0	  &1/2	   & 1     &0.0073(2) &0.0073(2) &0.0059(3) &0 &0 & 0.00365(9)\\
 Ta &2$b$           &0	   & 0	  &1/4	   & 1     &0.0039(1) &0.0039(1) &0.0046(1) &0 &0 &0.00195(4)\\
 Ta &6$h$           &0.49148(2)	&0.98296(3) &1/4	   & 1     &0.0056(1) &0.0039(1) &0.0041(1) &0 &0 &0.00193(4)\\
 Se &12$k$           &0.33781(3)	&0.16890(2)	&0.38160(2)	   & 1     &0.0045(1) &0.0045(1) &0.0046(1) &0.00022(3) &0.00043(6) &0.00226(5)\\
 Se &4$f$           &1/3	   & 2/3  &0.11041(3)      & 1     &0.0051(1) &0.0051(1) &0.0044(2) &0 &0 &0.00253(6)\\
\end{tabular}
\end{ruledtabular}
\end{table*}



First, the composition and structure of CVT-grown single crystals were solved by single crystal XRD analysis. Although the host TaSe$_2$ is known to crystallize in a variety of polytypes,\cite{Bjerkelund1967} we have determined that \cts\ only crystallized in the 2H-type structure, adopting the hexagonal centrosymmetric space group $P6_3$/$mmc$ (SG No. 194). The details of the structure solution are given in Table \ref{tab_scxrd}, and the refined structural parameters are summarized in Table \ref{tab_scxrd_r}. The occupancy parameter for all the elements were found to be unity, which results in stoichiometric \cts\ .

To check the consistency between different single crystals and their orientations, we examined the Laue diffraction patterns of several samples. The experimental Laue XRD image in Figure \ref{fig_xrd}(d) highlights the sixfold symmetry along the [001] direction which can be accurately simulated (Figure \ref{fig_xrd}(e)). The thin plate morphology therefore allows us to orient the crystal samples either in the $ab$-plane or along the $c$-axis for transport and magnetization studies.

The single crystal results also afforded satellite reflections that imply superlattice formation. TMDs of the 2H-type can adopt a 2$a_0$ $\times$ 2$a_0$ superlattice structure\cite{Xie2022} where $a_0$ is the unit cell parameter of the pristine 2H-TaSe$_2$ parent compound. Generally, this superlattice develops when $x$ = 1/4 in Nb and Ta-based TMDs. We find that \cts\ is isostructural to Ni$_{1/4}$TaSe$_2$ as well as several $M_{1/4}$TaS$_2$ TMDs.\cite{Maksimovic2022, Vanlaar1971, Morosan2007} Within the 2$a_0$ $\times$ 2$a_0$ superlattice, the intercalated cobalt occupies the 2$a$ Wyckoff position, leading to a 2D triangular sublattice of the Co$^{2+}$ cations. This triangular metal sublattice with Co--Co distances equal to the unit cell parameter $a$ is interesting in the context of magnetically frustrated systems.\cite{ANDERSON1973} It has also been implicated in forming toroidal structures\cite{Ding2021} in an ordered antiferromagnetic state. We discuss the potential implications of such a triangular sublattice later in this article.

For the polycrystalline samples, we confirmed their chemical stoichiometry via Rietveld analysis. A representative fit with a pXRD pattern is shown in Figure \ref{fig_xrd}(c). The lattice parameters from pXRD ($a$ = 6.8835(5) \AA\ and $c$ = 12.4791(8) \AA) at room temperature are very similar to those we found from the single crystal XRD at 250 K ($a$ = 6.88280(10) \AA\ and $c$ = 12.4535(3) \AA). 
The similarity of the unit cell dimensions indicates the closeness in composition and structure of our samples in both its polycrystalline and single crystal form.   

\subsection{Metal oxidation states and composition}

To better understand the implications of metal oxidation states on the magnetic and electronic properties, we analyzed some of the key x-ray photoelectron spectra (XPS) of the final product. First, we discuss tantalum. The asymmetric XPS peak shapes were fit using the Gelius profile convoluted with a Gaussian. Such asymmetry is characteristic of conducting samples. Figure \ref{fig_xps}(a) shows the Ta 4$f$ spin-orbit split peaks at 24.7 eV (4$f_{7/2}$) and 25.5 eV (4$f_{5/2}$) consistent with those previously observed in Fe$_{0.28}$TaSe$_2$.\cite{Chiang1976} Satellite peaks have been observed in TaSe$_2$\cite{Chiang1976, Pettenkofer1992} and are equally spaced by 1.9 eV as the main 4$f$ features. The additional peaks are inherent to samples with mixed oxidation states. From our fit, we find two distinguishable oxidation states, Ta$^{3+}$ and Ta$^{4+}$.

For the intercalated Co cations in the host, we find results largely consistent with a divalent metal. The binding energies at 777.9 and 793.1 eV in Figure \ref{fig_xps}(b) are close to those belonging to CoSe$_2$,\cite{Gao2018} while the broader features at 779.2 and 794.4 eV correspond to Co 2$p_{3/2}$ and 2$p_{1/2}$ observed in CoO respectively. The observation of satellite peaks separated by ~6.5 eV to the main Co 2$p_{3/2}$ confirms the Co$^{2+}$ oxidation state.\cite{BIESINGER2011} The presence of satellite peaks is characteristic of the high-spin state.\cite{PETITTO2008, Shen1990} However, due to the high content of surface oxidation, we cannot discard the possibility of these features being inherent to surface CoO itself. The binding energies for Se 3$d_{5/2}$ (53.8 eV) and 3$d_{3/2}$ (54.7 eV) are shown in the inset and are comparable to those observed in other similar transition metal diselenides.\cite{Chuang2020, Pettenkofer1992} Overall, the XPS results are consistent with Co$^{2+}$ coordinated to Se$^{2-}$ anions in the TMD. 

\subsection{Magnetic Susceptibility and magnetization}

\begin{figure}[!h]
\includegraphics[scale=0.395]{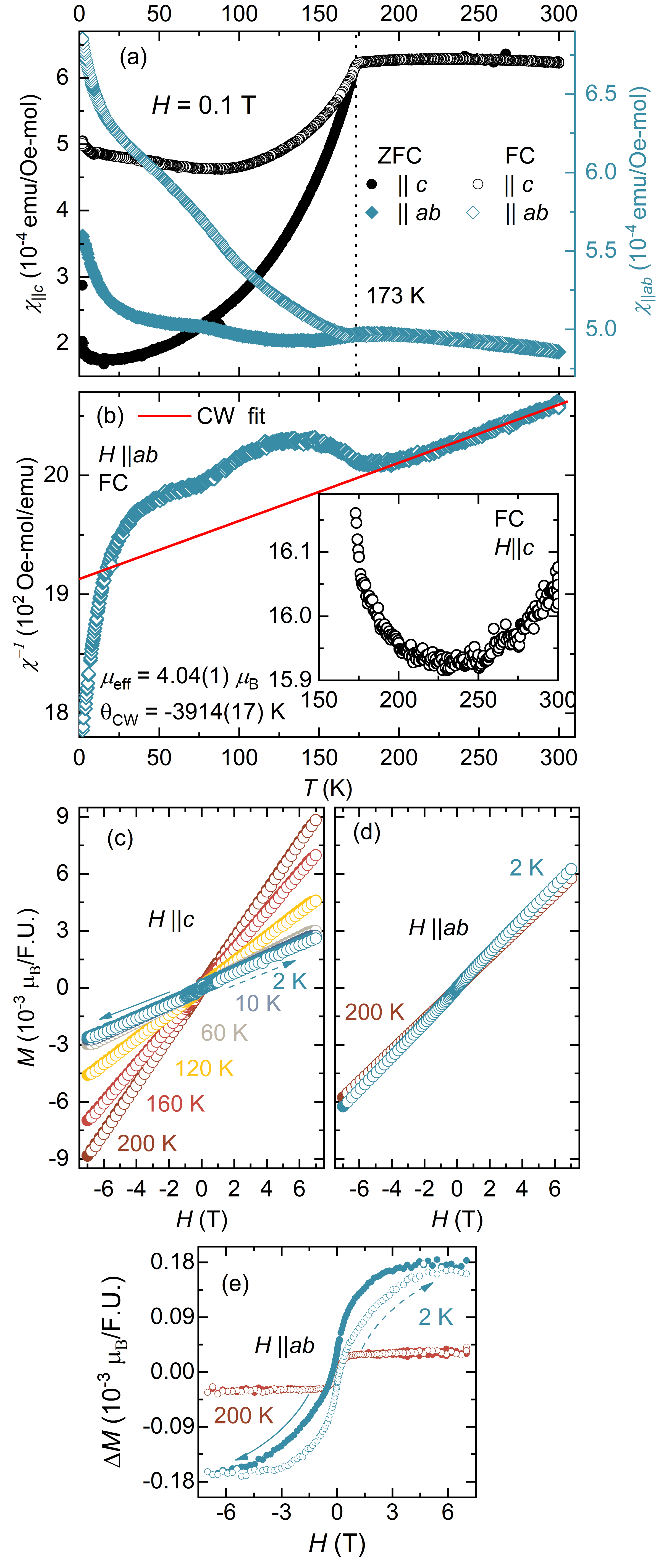}
\caption{\label{fig_mag} (a) Temperature-dependant ZFC and FC molar magnetic susceptibility with an applied field of $H$ = 1 kOe parallel to the $c$-axis and $ab$-plane.
(b) Inverse magnetic susceptibility with field applied in the $ab$-plane fitted to the Curie-Weiss law.
The inset shows the non-linearity of the inverse susceptibility with the field applied along the $c$-axis.
(c) Magnetization isotherms with field along the $c$-axis and (d) the $ab$-plane.
(e) FM component of (d) obtained by subtracting an AFM background slope.}
\end{figure}

The temperature-dependent magnetic susceptibilities ($\chi$) collected with fields applied along the $c$-axis and $ab$-plane are shown in Figure \ref{fig_mag}(a). An antiferromagnetic transition is observed at \TN\ = 173 K, and the clear difference between the susceptibilities $\chi_{||ab}$ and $\chi_{||c}$ indicates strong anisotropy below \TN . Such behavior suggests that the easy in \cts\ is along the $c$-axis. At room temperature $\chi_{||c}$ is greater than $\chi_{||ab}$, and this anisotropy resembles that of pristine 2H-TaSe$_2$\cite{BENCHIMOL1978}. Below \TN\, the susceptibility $\chi_{||ab}$ bifurcates between the ZFC and FC curves. Interestingly, the FC curve shows some degree of spin polarization along the $ab$-plane. The ZFC and FC splitting in $\chi_{||c}$ is also apparent, but the splitting is field-dependent and suppressed at higher appplied fields (see supplementary material).

We apply as a first approximation a Curie-Weiss fit to the linear portion of the inverse magnetic susceptibility (Figure \ref{fig_mag}(b)). The inset in Figure \ref{fig_mag}(b) shows the inverse susceptibility with the field applied out of plane. The curvature in the paramagnetic region suggests some local spin order above the \TN. Due to this high-temperature anomaly, we applied the Curie-Weiss law only to the inverse of $\chi||ab$. The negative Weiss temperature of $\Theta_{\mathrm{CW}}$ = - 3914(17) K indicates that the paramagnetic regime does not follow the Curie-Weiss law. Likely temperature-independent Pauli paramagnetism must also be added to correct for the abnormally large $\Theta_{\mathrm{CW}}$. Still, we extract a Curie constant of $C$ = 2.05(1) K emu/Oe-mol and an effective magnetic moment of \mueff\ = 4.05(1) \muB. This value is slightly greater than the calculated value for Co$^{2+}$ ($d^7$) in the high-spin state ($S = \frac{3}{2}$, \mueff\ =3.88 \muB). A high spin state for Co$^{2+}$ would be consistent with previously observed cobalt intercalated NbS$_2$ and TaS$_2$.\cite{VANDENBERG1968, Liu2022} A caveat here is that the heavier selenide may alter the ligand field strength compared to sulfide and therefore the spin-state of cobalt.

The magnetization vs. applied field results in Figure \ref{fig_mag}(c) and \ref{fig_mag}(d) demonstrate the strong anisotropy of the magnetic ordering. When applying the field out of the plane, the linear behavior of the magnetization displays a significant slope reduction below \TN. For all isotherms, the magnetization does not trace the same paths, likely caused by subtle hysteresis. The out-plane applied field measurement (Figure \ref{fig_mag}(c)) shows a reduction as a function of temperature as expected from AFM order. The in-plane measurement (Figure \ref{fig_mag}(d)), however, shows a slight increase in the magnetization at 2 K, suggesting potential spin canting along this direction. 

In order to reveal any subtle uncompensated magnetization within the $ab$-plane, we subtracted an antiferromagnetic component from the $M-H$ curves at 2 K and 200 K. As shown in c, blocking of the $M-H$ curve is apparent at 2 K, and the remnant magnetization at 2 K significantly increases compared to the 200 K. Switching the field direction reveals a soft hysteresis curve that saturates at $\Delta M$ = 0.000183 \muB\ per formula unit (Figure \ref{fig_mag}(e)). Due to the anisotropic character of this phenomenon, we attribute this observation to be intrinsic to the title compound. Such noncollinear antiferromagnetism has been observed before in related TMDs and implicated in the anomalous Hall effect.\cite{Ghimire2018, Tenasini2020, Park2022, Takagi2023} We explore such phenomena in \cts\ in the next section.


\subsection{Transport and magnetotransport properties}

\begin{figure}[t]
\includegraphics[scale=0.38]{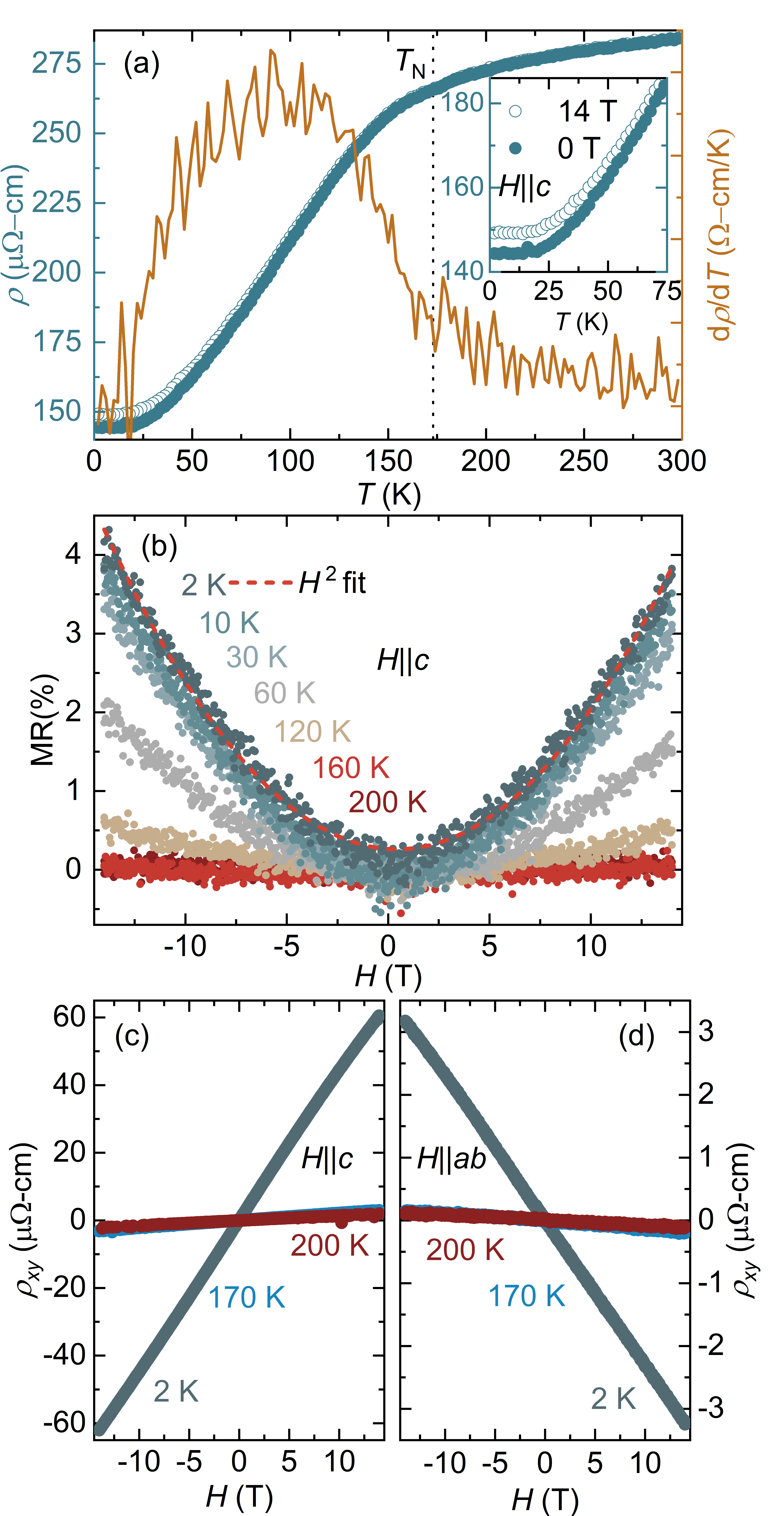}
\caption{\label{fig_R} Magnetotransport measurements of \cts. (a) Temperature dependence of the resistivity at 0 T and 14 T with current applied in the $ab$-plane. 
The inset highlights the low-temperature region of the resistivity. 
(b) Field-dependant magnetoresistance at various temperatures.
(c) Transverse resistivity showing Hall effect with field applied along the $c$-axis (c) and the $ab$-plane providing the planar Hall effect.}
\end{figure}

With the magnetic properties determined, we proceed to characterize the electrical transport properties of \cts. The low-temperature resistivity collected from 2 K to 300 K is shown in Figure \ref{fig_R}(a) at 0 T and 14 T applied field strengths. The resistance was measured in-plane with the magnetic field applied out of the plane. There are no noticeable differences upon measuring under ZFC or FC conditions. In the paramagnetic state at higher temperatures ($>$ 173 K), 
$\rho$($T$) has a linear trend and its slope is not altered immediately below \TN. Upon further cooling, however, the slope becomes noticeably steeper below 150 K, which is more noticeable in the derivative d$\rho$/d$T$ (Figure \ref{fig_R}(a)) Such a decrease in the resistivity is expected due to the reduction of electron scattering granted by long-range magnetic order.\cite{PARKIN1980, Parkin1980_2} The resistivity does not change much with an increase in the applied field to 14 T below \TN\ (inset of Figure \ref{fig_R}(a)). These magnetotransport results entail that long-range antiferromagnetism arises from electrons near the Fermi surface in \cts.

Past studies have demonstrated that the metallic character of Ta-based TMDs decreases when cations are intercalated into their vdW gaps.\cite{Bhoi2016, Liu2022} In the present case, the metallic character follows this same trend since \cts\ is less conductive than the pristine host\cite{Bhoi2016, Baek2022}. The residual resistivity ratio (RRR = $\rho_{\mathrm{300 K}}$/$\rho_{\mathrm{2 K}}$) of our single crystal sample is 1.97 and its residual resistivity ($\rho_0$) is  144.3 $\mu\Omega$-cm. 

Iwaya \textit{et al.,} have noted that the magnitude of the RRR in 2H-NbSe$_2$ correlates to the prominence of the CDW ordering transition.\cite{Iwaya2003} Samples with RRR$>$10 exhibit a stronger CDW appearance during resistivity measurements. This value can be used as an indirect way to estimate the completeness of the CDW order. At $T<$\TN, the resistivity contributions should be minimized providing a higher RRR in \cts. However, if spin fluctuations persist down to low temperatures in \cts\, such contributions to the resistivity remain present. Overall, the low RRR in \cts\ below the AFM transition likely makes the emergence of CDW unlikely.

Further exploring the field dependence of the resistivity, we measured magnetoresistance (MR) from 2 K to 200 K and up to an external field strength of 14 T (Figure \ref{fig_R}(b)). The MR curves obtained as MR = [$\rho$($H$)-$\rho$(0)/$\rho$(0)]$\times$100 show an appreciable change only below 120 K instead of 160 K and below. This surprising result in the MR change stands in distinction to the $M$ vs. $H$ curves (Figure \ref{fig_mag}(c)) where a response or change in slope was more perceptible closer to the \TN \, of 173 K.

 According to Kohler's rule, the MR should vary proportionally to the square of the applied magnetic field multiplied by the scattering time for a conventional metal. By plotting the MR vs. $H/\rho_0$, we did not observe this field dependency in the present sample (see supplementary information). Hence, we attribute the increase in MR effect to long-range magnetic order where the MR reaches ~4\% at 2 K with no significant changes up to 30 K. 

The MR response in \cts\ is notably different from that of the isostructural Fe$_{1/4}$TaSe$_2$ and Ni$_{1/4}$TaSe$_2$, which are both ferromagnets.\cite{Maksimovic2022, Nosova2024} Unlike Fe$_{1/4}$TaS$_2$\cite{Morosan2007} and Fe$_{1/4}$TaSe$_2$\cite{Nosova2024}, the MR in \cts\ never exhibits any switching in sign of the MR upon reaching a critical field strength. Instead, it displays a quadratic response despite the field being aligned along the easy axis up to 14 T. The contrasting behavior between the Fe, Co, and Ni analogs is direct evidence of the influence that the intercalated magnetic ion has over the exchange interactions affecting long-range magnetic order and consequently the electronic behavior.

We now present our exploration of whether \cts\ expresses the anomalous Hall effect as found in other TMD systems. The transverse resistance in Figure \ref{fig_R}(a) shows the Hall effect measurements with the field parallel to the $c$-axis. The anomalous Hall effect can be expressed as $\rho_{xy}$ = $R_{H}(B)$ + $R_A(M)$ in ferromagnetic materials, where the first term denotes the ordinary Hall effect caused by the Lorentz force acting on the charge carriers; the second term expresses the anomalous contributions that generally originate from the internal magnetic order. Although there is a minimal linear deviation that may be caused by the interaction of the carriers and the magnetic order, we do not find a significant anomalous contribution.

The Hall effect measurements do yield other important aspects of \cts. The positive slope of $\rho_{xy}$ signifies hole-type carriers. We obtain the charge carrier density with $n$ = 1/$R_{H}$, resulting in $n_{2K}$ $\approx$ 1.4 $\times$10$^{22}$/cm$^3$, $n_{170K}$ $\approx$ 2.8 $\times$10$^{23}$/cm$^3$, and $n_{200K}$ $\approx$ 4.0 $\times$10$^{23}$/cm$^3$. The decrease in carrier density upon cooling suggests a reduction in the Fermi surface induced by the development of antiferromagnetic order. 

Although we did not find an anomalous contribution to the Hall effect, we must issue a caveat on the limitations of our measurement. The easy axis is along $c$, and we encountered sample geometry limitations to measure the conventional Hall effect along this easy axis. We thus measured the planar Hall effect with current perpendicular to the field as shown in Figure \ref{fig_R}(d). Overall, the present sample does not display any additional contributions from the ordinary Hall effect in this configuration. Other intercalated TMDs such as Mn$_{1/4}$NbS$_2$, however, afford complex anomalous Hall effect contributions below the transition temperature with in-plane ferromagnetic moments\cite{Polesya2020}. 

\subsection{Specific heat and magnetic entropy analysis}

\begin{figure}[t]
\includegraphics[scale=0.35]{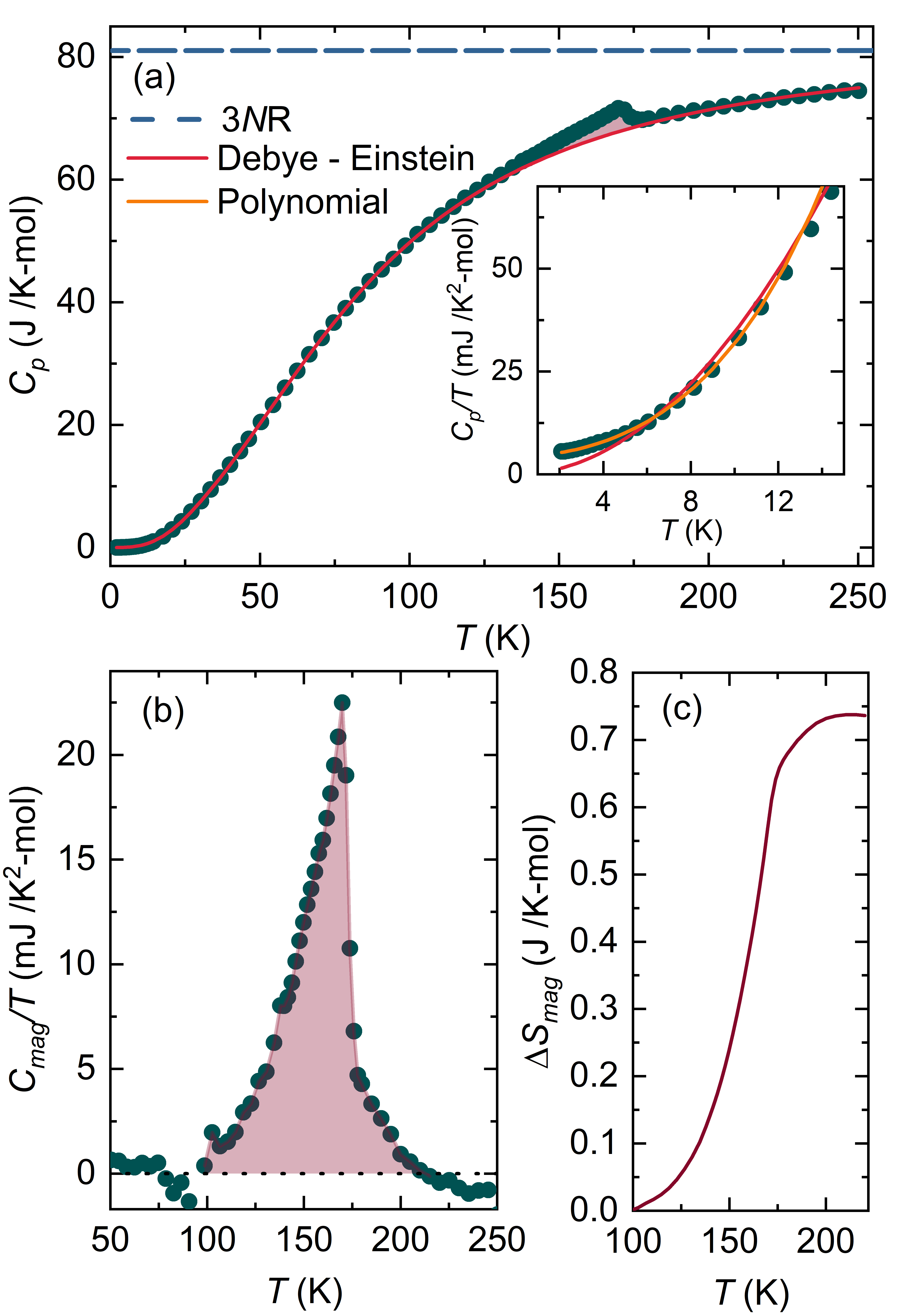}
\caption{\label{fig_Cp} Specific heat measurement of \cts. (a) Specific heat fitted to the Debye-Einstein model and a general polynomial
in the low-temperature region (inset). 
The inset is plotted in $C_p/T$ showing the deviation form both models.
(b) Magnetic specific heat. The shaded area indicates the integrated region used to obtain (c) the magnetic entropy from 100 K to 220 K.}
\end{figure}

To better characterize the contributions of the AFM order on the specific heat of \cts, we analyze the electronic and phonon contributions as well. The temperature dependence of the specific heat $C_{\mathrm{p}}$($T$) is shown in Figure \ref{fig_Cp}(a). The high-temperature region of the $C_{\mathrm{p}}$($T$) approaches the classical Dulong-Petite value 3$N$R = 81 J mol$^{-1}$K$^{-1}$, where R is the ideal gas constant (8.314 mol$^{-1}$K$^{-1}$) and $N$ the number of atoms. A characteristic second-order transition peak is observed at \TN\ $\approx$ 173 K belonging to the emergence of the AFM order. 

The inset in Figure \ref{fig_Cp}(a) shows the low-temperature region of the specific heat and a fit to a general polynomial function to obtain the electronic and vibration contributions. The crossover temperature where the electronic specific heat becomes greater than that of the lattice is seen at ~5 K. Therefore, the low-temperature region from 2 K to 12 K was fit with $C_{\mathrm{p}}$/$T$ = $\gamma$ + $\beta$$T^2$ + $\delta$$T^4$, where the electronic contributions are given by the Sommerfeld coefficient $\gamma$ = 4.38(8) mJ mol$^{-1}$K$^{-2}$. The resultant $\gamma$ is comparable to that of the host 2$H$-TaSe$_2$
and that of other intercalated TMDs.\cite{Li2017, Bhoi2016,Li2010} The Debye temperature (\DT) in this region can be obtained from $\beta$ =  0.21(3) mJ mol$^{-1}$K$^{-4}$ by using \DT$^{low-T}$ = (12$\pi^4N$R/5$\beta$)$^{1/3}$. The resulting \DT$^{low-T}$ = 310(1) K, although larger than in the parent compound, is similar to the sulfide counterpart.\cite{Liu2022} Such contrast may be a consequence of the symmetry changes caused by the increasing amount of intercalated cation amount, $x$.\cite{Bhoi2016, Naik2022, Luo2017} The $\delta$ term is added to account for the fact that the $T^3$-law is only valid through \DT/50 = 6.2 K\cite{Tari2003} resulting in $\delta$ = 6.2(2)$\times$10$^{-4}$ mJ mol$^{-1}$K$^{-6}$.

\begin{figure*}
\includegraphics[scale=0.36]{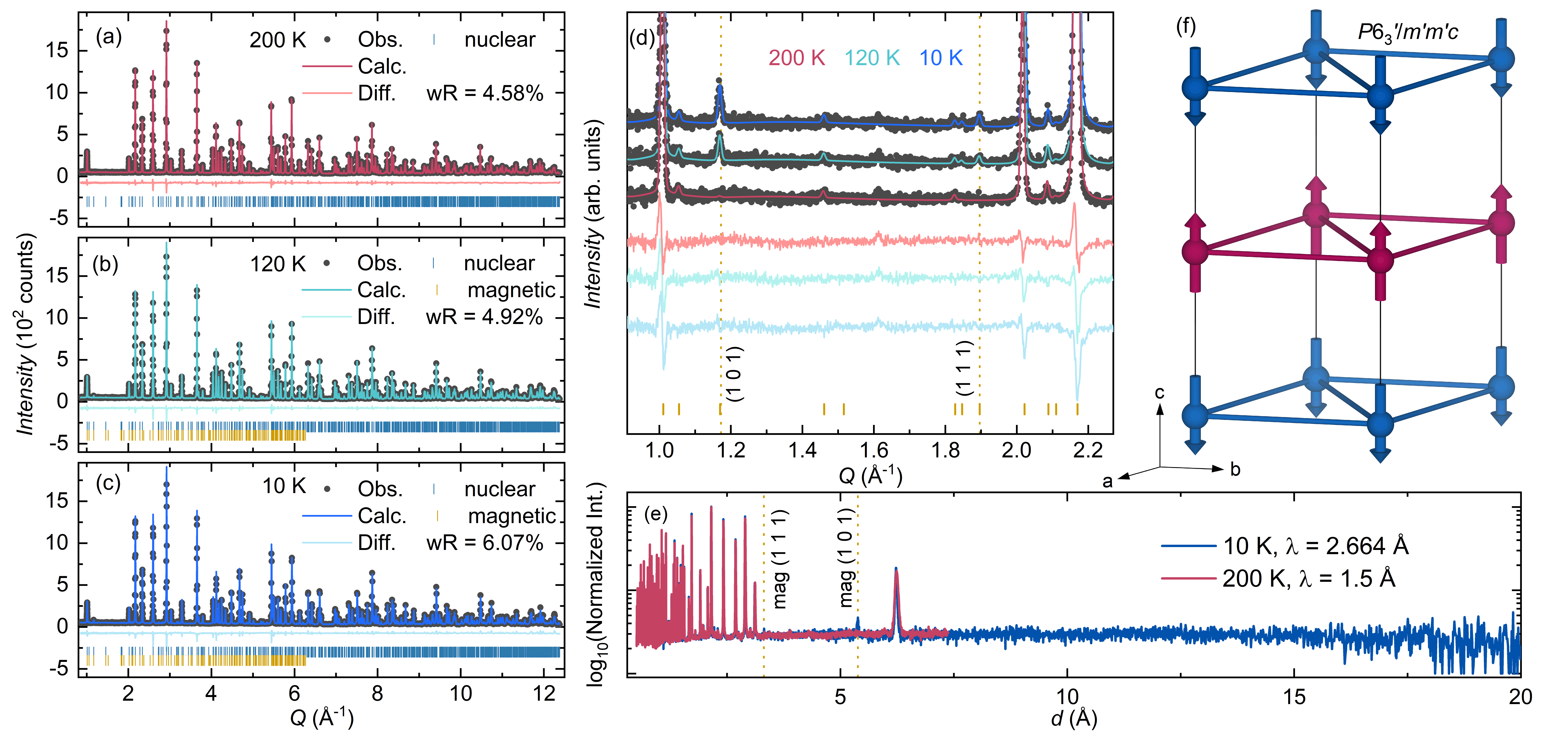}
\caption{\label{fig_NPD} Neutron powder diffraction of \cts. The data is fit using the Rietveld method at (a) 200 K, (b) 120 K, and (c) 10 K. Only a magnetic phase is added for the 120 K and 10 K datasets.
(d) Low-$Q$ region of (a-c) highlighting the emerging magnetic reflections corresponding to (101) and (111).
(e) absence of magnetic reflections at higher $d$-spacing.
(f) A-type antiferromagnetic structure of \cts\ with propagation vector $k$ = (0 0 0).}
\end{figure*}

To inspect the magnetic specific heat $C_{\mathrm{mag}}$, the high-temperature lattice contributions were subtracted from the total $C_{\mathrm{p}}$($T$). We consider $C_{\mathrm{lattice}}$ = $C_{\mathrm{Debye}}$ +$C_{\mathrm{Einstein}}$, where:

\begin{eqnarray}
    C_{\mathrm{Debye}} = 9n\mathrm{R}\left( \frac{T}{\Theta_\mathrm{D}}\right)^3\int_{0}^{\Theta_\mathrm{D}/T} \frac{x^4e^x}{(e^x-1)^2}\,dx
\label{eq_Debye},
\end{eqnarray}
and
\begin{eqnarray}
    C_{\mathrm{Einstein}} = 3a\mathrm{R}\left(\frac{\Theta_\mathrm{E}}{T}\right)^2\frac{e^{\Theta_\mathrm{E}/T}}{(e^{\Theta_\mathrm{E}/T}-1)^2} 
\label{EinsteinCp},
\end{eqnarray}

The best fit for the full temperature range was obtained with one Debye term giving \DT\ = 190(5) K, $n$ = 1.22(15), and an Einstein term \ET\ = 313(4) K, $a$ = 2.07(6), where $N$ = $n$ + $a$ accounts for the total number of atoms per formula unit. The magnetic contribution to the specific heat isolated by subtracting the Einstein-Debye model is shown in Figure \ref{fig_Cp}(b). The magnetic entropy in Figure \ref{fig_Cp}(c) was then approximated by integrating $\Delta$$S_{mag}$ = \(\int_{}^{} C_p(T)/T \,dT\) yielding $\sim$26\%\ of the expected value for $S$ = 3/2. This lower than expected value may indicate that the Co$^{2+}$ cation may either not be in a high-spin configuration, or that significant magnetic fluctuations persist below the ordering temperature due to the triangular lattice. If the former, then the $d^7$ cation in a low-spin state would yield a magnetic entropy term better described by $S$ = 1/2. However, as we describe below in our neutron results, this low-spin argument may not be the most probable given that this diminished moment size has been observed in other related TMDs.

\subsection{Neutron powder diffraction: Collinear antiferromagnetism}

To elucidate the long-range magnetic structure, we conducted neutron powder diffraction analysis above and below \TN, specifically at 200 K, 120 K, and 10 K (Figure \ref{fig_NPD}(a-c)). Based on the results from magnetic susceptibility and specific heat measurements, we observed a magnetic order transition at \TN\ = 173 K. The emergence of magnetic Bragg reflections in the NPD patterns is discernible at 120 K and 10 K as highlighted by the dashed line in Figure \ref{fig_NPD}(d). Two configurations of the time-of-flight diffractometer were used to cover the $d$-range from 0.485 \AA\ to 13 \AA\ and 1.07 \AA\ to 21 \AA\ (Figure \ref{fig_NPD}(e)). This wide $d$-range was to ensure the observation of all possible magnetic reflections. In both configurations we observed only two magnetic reflections corresponding to the (101) and (111) planes. Only the former coincides with a nuclear Bragg index as the (111) is disallowed by the systematic absences of the parent space group. Nevertheless, the magnetic unit cell coincides with that of the nuclear unit cell, and thus the magnetic propagation vector \textbf{$k$} = (0, 0, 0). 

We solved the magnetic structure in our Rietveld refinements with the NPD data by analyzing the possible magnetic symmetries provided by k-SUBGROUPGMAG\cite{k-subgroupmag} tool from the Bilbao Crystallographic Server within GSAS-II. Of the 8 possible magnetic subgroups, the best refinement was obtained with $P$6$_\mathrm{3}'\slash m'm'c$ magnetic space group (No. 194.268). This magnetic order is isostrutural to the sulfide-based antiferromagnets Cr$_{1/4}$NbS$_2$ and Fe$_{1/4}$NbS$_2$\cite{Vanlaar1971, Lawrence2023}. Such magnetic symmetry allows for ferromagnetic moments aligned in the $ab$-plane while antiferromagnetically along the $c$-axis (Figure \ref{fig_NPD}(f)). The moment direction is collinear and fully along the $c$-direction, consistent with the magnetic susceptibility and magnetization results in Figure \ref{fig_mag}. This configuration is an $A$-type antiferromagnet with all moments collinear to the [001] direction.

The total magnetic moment obtained from neutron diffraction at 120 K is 1.08(11) \muB\ and increases to 1.35(11) \muB\ at 10 K per Co site. These results indicate that there is missing moment if Co$^{2+}$ in the high-spin state assuming that $g$ = 2 for $\mu$ = $g\cdot S$. Consequently the observed total moment is about a third of the expected 3 \muB \, for the high-spin state. Refinement of the magnetic moment in the Ta site was attempted, but the resulting moment was negligibly small and thus fixed to zero. Such phenomena where the magnetic moment is smaller than expected for the intercalated magnetic ion has been observed in other intercalated tantalum-based TMDs,\cite{Parkin1983, Park2023, An2023} but not to the extent seen in our results. Our results in moment size match that of the sulfide analogue Co$_{1/3}$TaS$_2$,\cite{Park2023} where the authors also recover about a third of the expected moment. This reduction in the ordered moment for Co-intercalated Ta-based TMDS could be attributed to the metallic nature of these compounds. The interaction between a localized band due to Co and an itinerant $5d$-band due to Ta could lead to partial screening of the local moments. As mentioned earlier in the discussion of the specific heat measurements, another possible explanation could include quantum spin fluctuations due to the frustrated triangular sublattice of the Co cations. Both explanations are worth pursuing with more studies that include inelastic neutron scattering. 

Finally, the temperature-dependent NPD patterns allowed us to search for possible CDW formation in \cts. At all temperatures, no indication of CDW ordering was found as known to occur in the parent 2H-TaSe$_2$\cite{Moncton1975}. The nuclear structure could be fully fit with the centrosymmetric hexagonal structure ($P6_3$/$mmc$), consistent with the single crystal XRD results. The refined lattice parameters from neutron diffraction are $a$ = 6.88155(7) \AA\ and $c$ = 12.45907(8) \AA\ at 200 K decreasing to $a$ = 6.87637(8) \AA\ and $c$ = 12.43289(9) \AA\ at 10 K. Notably, the Co-Co distance is shorter along the $c$ direction than along the interplanar direction. This interplanar distance is further shortened at lower temperatures. The contraction reduces the Co-Ta distance by 0.0065(4) \AA\ from 200 K to 10 K. This reduction in proximity could have implications on the orbital overlap between metal centers. We now discuss the electronic structure and how it could be responsible for an itinerant AFM ground state in \cts.


\section{Itinerant A-type AFM origins}

Based on the smaller than expected value of the ordered magnetic moment from NPD and specific heat measurements, we can surmise that the emergent magnetic order is tied to the itinerant nature of \cts. The absence of a substantial amount of magnetic entropy from Figure \ref{fig_Cp}(c) contributes to the idea that the 3$d$ electrons of cobalt should not be considered entirely localized\cite{Clinton1975}. It is understood that the intercalated ions strongly donate electrons into the Ta 5$d_{z^{2}}$ conduction band\cite{DiSalvo1973, Motizuki1992, PARKIN1980_3}.
Furthermore, when the intercalated ion is a first-row transition metal, its own $d$-orbitals hybridize substantially with the Ta $d_{z^2}$ band. In the case of Co$^{2+}$, we find an expected \mueff\ close to that of $S$ = 3/2 but the abnormally large  $\mid\Theta_{\mathrm{CW}}\mid$ points to the $\chi$ having a large Pauli paramagnetic contribution, which further supports the notion of delocalized magnetism. 

Since \cts\ and its sulfide analog are structurally similar, the main exchange pathways are expected to be similar. Their energies, however, are obviously different enough to lend to two distinct ground states. 
While the sulfur counterpart has been reported to present ferromagnetic interactions\cite{Liu2022, Liu2021}, we have observed the present composition to have an AFM order with an easy axis along the [001] direction. 
A key reason to this distinction may be the non-centrosymmetric spacegroup in which Co$_{0.22}$TaS$_2$ crystallizes leading to various types of coupling, most importantly the Dzyaloshinskii–Moriya interaction\cite{Xie2022}. 
Provided that the 5$d_{z^{2}}$ orbital is effectively hybridized with the Co 3$d$ in \cts, it would imply a stronger one-dimensional character propagating in the vertical direction reducing the dimensionality of the strong interactions.

\begin{figure}
\includegraphics[scale=0.31]{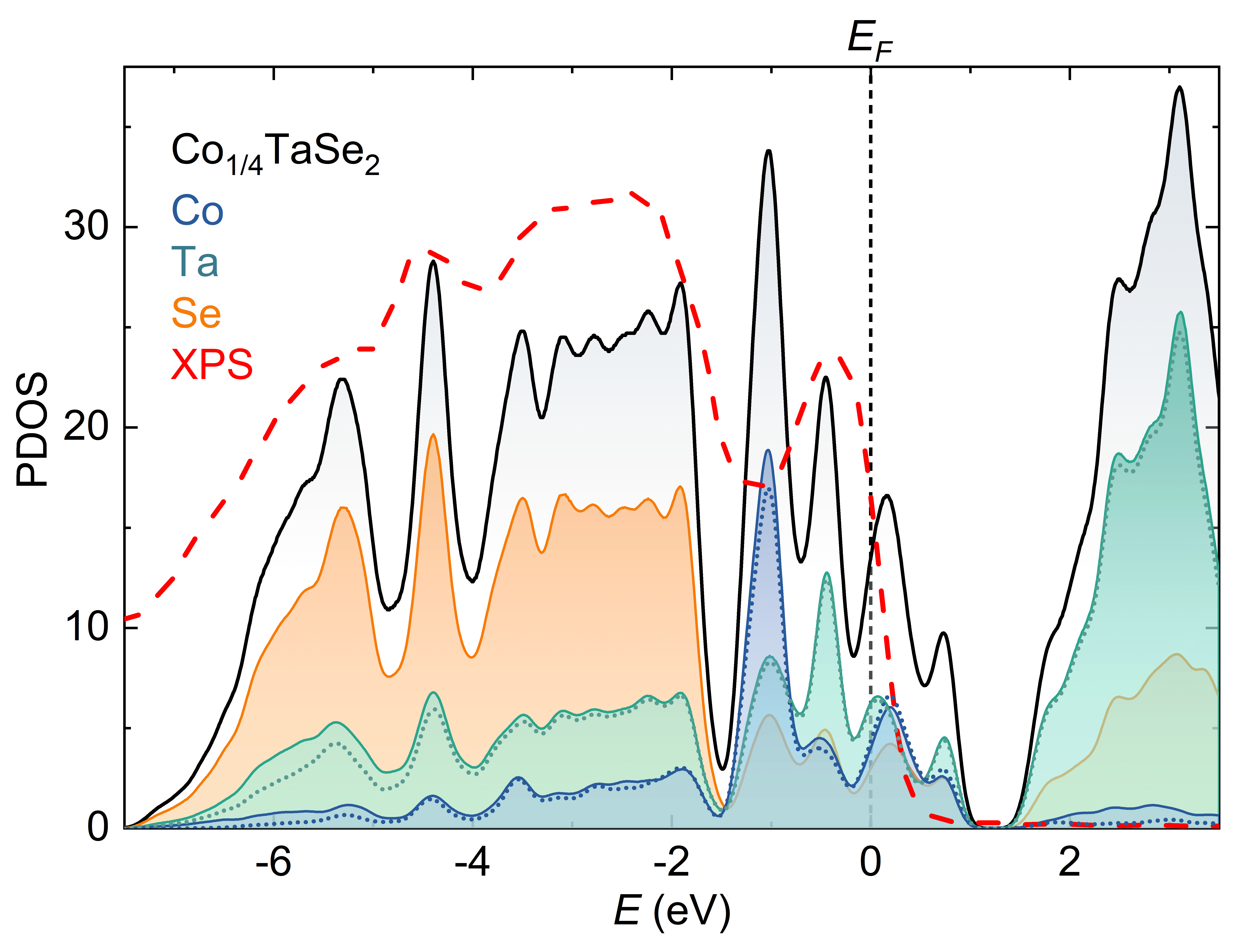}
\caption{\label{fig_PDOS} Non magnetic partial density of states. The black curve shows the total density of states of \cts\ which resembles that of the experimentally observed states through XPS (red dashed curve). The dotted lines indicate the contributions from the Co 3$d$ (blue) and Ta 5$d$ orbitals (green). The overestimated bands at -1.0 eV belong primarily to all the Co 3$d$ orbitals except $d_{z^2}$. }
\end{figure}

The partial density of states calculation in Figure \ref{fig_PDOS} corroborates these arguments and is comparable with the experimental XPS near the Fermi edge. The valance bands close to the Fermi level (-0.44 eV) are predominantly those belonging to the $d$ orbitals of Co and $d_{z^{2}}$ of Ta. The Co and Ta vertical arrangement supports the argument that the exchange interaction between the Co magnetic ions is mediated along the vertically orientated $d_{z^{2}}$ orbital\cite{Zhao20171}. This band overlap is likely strengthened as a function of lattice contraction observed in comparing the $c$ lattice parameter from neutron diffraction at high (200 K) vs low temperatures (10 K).

In the case of the sulfur analog, the ferromagnetic exchange mechanism is of the RKKY interaction type\cite{Liu2021}. Such interaction is mediated by the conduction electrons\cite{Park2023} and it is therefore equally likely to fit the present case. Here we find a ferromagnetic coupling in plane and antiferromagnetic interaction out of plane. Selenium thus plays an important role in the exchange mechanism.

 What is interesting to consider is why \cts\ is an A-type itinerant antiferromagnet while Fe$_{0.25}$TaSe$_2$ and Ni$_{0.25}$TaSe$_2$ are ferromagnets. In the RKKY mechanism, the distance variable in energy expression determines the nature of the magnetic ground state\cite{Kasuya1956, Ruderman1954, Xie2022}. Within the intercalated TaSe$_2$ layers, the $a$-axis remains largely unaffected ($\sim$6.88 \AA). Therefore, the fixed intraplanar distance leads to ferromagnetic coupling within the triangular sublattice regardless of whether the intercalated ion is Fe, Co, or Ni. However, comparing the $c$-axis of Co$_{0.25}$TaSe$_2$ ($c$ = 12.4791(8) \AA) to that of the ferromagnets Fe$_{0.25}$TaSe$_2$ ($c$ = 12.7394(2)) \AA) and Ni$_{0.25}$TaSe$_2$ ($c$ = 12.506(5) \AA), we find that the \textit{interplaner} metal-metal distance is the shortest for the present compound. The out of plane Co--Co distances is below a critical distance to flip the exchange coupling from ferromagnetic to an antiferromagnetic along the $c$-direction. Despite its different ground state compared to closely related compounds, \cts exhibits magnetic behavior consistent with other metallic TMDs whereby the conduction electrons interact strongly with the isolated spins and mediate exchange mechanisms.\cite{Liu2022, An2023, Park2023, Parkin1980_1}

Finally, the long-range magnetic order does not seem to be related to the CDW that the vacant host lattice presents. Intercalating metal cations into the vdW gap of the host TMD disrupts the charge ordering likely formed from lattice distortions rather than Fermi surface nesting\cite{Laverock2013, Harper1977, Baek2022}. We do not, however, discard the possibility of such phenomena to be found in an interplay with the magnetic ions. We speculate this is unlikely as the long-range magnetic order emerges at a higher temperature than the CDW suppressing its emergence.


\section{Conclusion}
In summary, we have grown single crystals of 2H-\cts\ and characterized the physical and magnetic properties of this novel intercalated transition metal dichalcogenide. Our measurements and analysis reveal the antiferromagnetic nature of \cts\ with collinear $A$-type antiferromagnetic order along the $c$-axis. We observe that there is a significant deficit of magnetic moment and entropy of the localized Co 3$d$ orbitals. This is best explained via an interplay between cobalt's 3$d$ orbitals and the itinerant electrons of Ta ions, which dominates the interaction with neighboring Co captured by our electronic density of states calculations. Deeper theoretical work is needed to provide insight into the origin of the itinerant antiferromagnetism. 
Our study on the Hall effect measurements reveals an ordinary behavior inherent to a classical metal expected from a traditional collinear AFM.

Since \cts\ is a new itinerant antiferromagnetic, we checked its spin order to see if it qualifies as a potential altermagnet. The nuclear symmetry in conjunction with the magnetic structure obtained from NPD was used in the program Amcheck\cite{Amcheck2024}, which is a symmetry analysis program to search for potential altermagnets.  The Amcheck program indicates that \cts\ is an altermagnet candidate. The nuclear and magnetic symmetry, as well as the observation of an in-plane ferromagnetic component, suggest that the title compound be further investigated as an altermagnet. 

\section{Acknowledgments}

The authors thank the U.S. Department of Energy
(DOE), Office of Science (Grant DE-SC0016434), for financial support.
Use of the Spallation Neutron Source (POWGEN experiment IPTS-31862) at Oak Ridge National Laboratory was supported by the U. S. Department of Energy, Office of Science, Office 
of Basic Energy Sciences, under Contract No. DE-AC02-06CH11357.
Measurements were performed with the support of the National Science Foundation (NSF-DMR 2105191).
We thank Harikrishan S. Nair for the discussions.
We also acknowledge support from the Quantum Materials Center. 
This work has been supported in part by the X-ray Crystallographic Center at 
The University of Maryland.

\nocite{*}

\bibliography{Co14TaSe2_Ref}

\end{document}